\documentclass[twocolumn,showpacs,preprintnumbers,amsmath,amssymb]{revtex4}

\usepackage{units}
\usepackage{graphicx}
\usepackage{dcolumn}
\usepackage{bm}
\usepackage[usenames]{color}
\definecolor{Blue}{rgb}{0,0.08,0.65}
\definecolor{Red}{rgb}{0.65,0.08,0.05}
\definecolor{Green}{rgb}{0.15,0.45,0.25}

\global\long\def\vx{{\bf x}}

\begin{document}


\title{ The invariant joint distribution of a stationary random field and its derivatives:\\
Euler characteristic and critical point counts in 2 and 3D}

\author{Dmitry Pogosyan${}^{1}$ }
\author{Christophe Gay${}^{2}$ }
\author{Christophe Pichon${}^{2}$ }%
\affiliation{%
${}^{1}$ Department of Physics, University of Alberta, 11322-89 Avenue, Edmonton, Alberta, T6G 2G7, Canada\\
${}^{2}$ Institut d'astrophysique
de Paris, 
98, bis boulevard Arago, 75 014, Paris,
France
}

\date{\today}

\begin{abstract}
The full moments expansion of the joint probability distribution
of an isotropic random field, its gradient and invariants of the Hessian
is presented  in 2 and 3D.
It allows for explicit expression for the Euler characteristic in ND and 
computation of extrema counts as functions of the excursion set threshold
and the spectral parameter, as illustrated on model examples.
\end{abstract}


\maketitle

Random fields are ubiquitous phenomena in physics appearing in areas from
turbulence to the landscape of string theories.
In cosmology, 
the large scale distribution of matter (LSS) and
the sky-maps of the polarized Cosmic Microwave Background (CMB) radiation 
-- two focal topics of current research --
are described as, respectively, 3D and 2D random fields.  
Modern view of the Universe, developed primarily through statistical
analysis of these fields, points to a Universe that is 
statistically homogeneous and isotropic with a hierarchy of structures 
arising from small Gaussian fluctuations of quantum origin.

While the Gaussian limit provides the fundamental starting point in the study
of random fields \cite{Adler,Doroshkevich,BBKS},
non-Gaussian features of the CMB and LSS fields are of great interest.
CMB inherits high level of Gaussianity from the
initial fluctuations, and small non-Gaussian deviations may
provide a unique window into the details of processes in the early Universe.
The gravitational instability that nonlinearly maps the initial Gaussian 
inhomogeneities in matter density into the LSS, on the other hand,
induces strong non-Gaussian features culminating in the formation of
collapsed, self-gravitating objects such as galaxies and clusters of galaxies.
At supercluster scales where non-linearity is mild, the non-Gaussianity of
the matter density is also mild, but still essential for quantitative 
understanding of the filamentary Cosmic Web \cite{bkp}
in-between the galaxy clusters.

The search for the best methods analyzing 
non-Gaussian random fields both in weak and strong
regimes is ongoing.  We focus on the
statistics of geometrical and topological properties of the
field that includes the Euler characteristic of excursion sets 
\cite{Gott1988,Matsubara0,Matsubara,2000PhRvL..85.5515C}
(and the rest of Minkowski functionals \cite{Buchert}),
the density of extremal points and statistics of
critical lines, or the skeleton \cite{NCD,pogoskel}. 
In this paper we present the formalism for 
computing such geometrical statistics for the homogeneous and isotropic
mildly non-Gaussian fields that can be represented or approximated by
a Gram-Charlier expansion around the Gaussian limit, to an arbitrary order
of the expansion. 

A statistically homogeneous ND random field $x$
is fully characterized by the
joint one-point distribution function (JPDF) of its value and its derivatives 
$P(\vx)$, $\vx\equiv (x,x_i,x_{ij},x_{ijk},\ldots)$.
We consider non-Gaussian fields $P(\vx)$ 
represented by the Gram-Charlier expansion \cite{Chambers}
\begin{equation}
P(\vx)=G(\vx) \left[1+\sum_{n=3}^\infty \frac{1}{n!}\;
\mathrm{Tr}\left[ \langle {\mathbf x}^n\rangle_{\scriptscriptstyle{\mathrm{GC}}} \cdot {\mathbf h}_n(\vx)\right]
\right]
\label{eq:defGP}
\end{equation}
around the Gaussian 
$G(\vx)\equiv(2\pi)^{-N/2}|{\bf C}|^{-1/2}\exp(-\frac{\scriptstyle 1}{\scriptstyle 2} \vx \cdot {\bf C}^{-1} \cdot \vx)$
that is arranged to match the mean and the covariance
${\bf C}=\langle \vx \otimes \vx\rangle$ of  the $\vx$ variables.
We consider all the variables to be defined as having zero mean
and unit variance, ${\bf C}_{ii}=1$.
The correction to the Gaussian approximation is the series in Hermite tensors
$ {\bf h}_n(\vx)=(-1)^{n}G^{-1}(\vx)\partial^n G(\vx) /\partial \vx^n$
of rank $n$ with coefficients  constructed from the moments, 
$\langle {\vx}^n\rangle_{\scriptscriptstyle{\mathrm{GC}}}=\langle \mathbf{h}_n(\vx) \rangle $.

The statistics of Euler characteristic and extremal points  \cite{Adler}, and
the basic description of the critical lines of the field in
the stiff approximation
\cite{pogoskel}, require to know the JPDF only up to
the second derivatives. In the Gaussian limit, in this case, the only 
non-trivial covariance parameter is the cross-correlation between the field and
the trace of the Hessian $\gamma=- \left\langle x \mathrm{Tr}(x_{ij}) \right\rangle$ 
\citep{BBKS,pogoskel}.

The JPDF in the form of Eq.~(\ref{eq:defGP}) is not ideal to
study critical sets statistics since the coordinate representation 
masks the isotropic nature of the statistical descriptors.
The pioneering works \cite{Matsubara0,Matsubara},
where the first correction to Euler characteristic was computed, 
demonstrate the arising complexities.
Instead we develop the equivalent of the Gram-Charlier expansion
for the JPDF of the field variables that are invariant
under coordinate rotation. Such distribution can be computed via 
explicit integration of the series Eq.~(\ref{eq:defGP})
over rotations, however we obtain it directly from general principles:
{\it the moment expansion of the non-Gaussian JPDF corresponds
to the expansion in the set of polynomials which are orthogonal with respect
to the weight provided by the JPDF in
the Gaussian limit}. Thus, the problem is reduced to finding such polynomials
for a suitable set of invariant variables.

The rotational invariants that are present in the problem 
are: the field value $x$ itself, the modulus of its gradient, 
$q^2=\sum_{i} {x_i}^2$ and the invariants of the matrix
of the second derivatives $x_{ij}$.
A rank N symmetric matrix has N invariants with respect to rotations.
The eigenvalues $\lambda_i$ provide one such
representation of invariants, however
they are complex algebraic functions of the matrix components. An alternative 
representation is a set of invariants that are polynomial in $x_{ij}$,
with one independent invariant polynomial per order, from one to N. 
A familiar example is the set of coefficients, $I_s$, of the characteristic equation for the eigenvalues, where the linear invariant 
is the trace, $I_1=\sum_{i} \lambda_i$, 
the quadratic one is $I_2=\sum_{i<j} \lambda_i \lambda_j$
and the N-th order invariant is the determinant of the matrix.
$I_N=\prod_{i} \lambda_i $.
Aiming at simplifying
the JPDF in the Gaussian limit 
\footnote{
By Gaussian JPDF we mean
the limit of the Gaussian field. Note that even in this limit, the
rotation invariant variables themselves are not, in general, Gaussian.
}.  
(e.g., \cite{Doroshkevich,pogoskel}, Appendix A) we use their linear 
combinations $J_s$
\begin{equation}
J_1 = I_1 ~,\quad 
J_{s\ge 2}  = I_1^s - \sum_{p=2}^s \frac{(-N)^p C_{s}^p}{(s-1)C_N^p} I_1^{s-p} I_{p}
\label{eq:Js}
\end{equation}
where $J_{s\ge2}$ are (renormalized) coefficients of the characterstic
equation of the {\it traceless part} of the Hessian and are independent
in the Gaussian limit on the trace $J_1$. 

Let us consider the 2D and 3D cases explicitly.
Introducing $\zeta=(x + \gamma J_1)/\sqrt{1-\gamma^2}$ in place of the
field value $x$ we find that the 2D Gaussian JPDF 
$G_{\rm 2D}(\zeta ,q^2,J_1,J_2)$,
normalized over $d\zeta dq^2 dJ_1 dJ_2$,
has a fully factorized form
\begin{equation}
G_{\rm 2D} = \frac{1}{2 \pi} 
\exp\left[-\frac{1}{2} \zeta^2 - q^2 - \frac{1}{2} J_1^2 - J_2 \right] \,.
\label{eq:2DG}
\end{equation}
Used as a kernel for the polynomial expansion, $G_{\rm 2D}$
leads to a non-Gaussian rotation invariant
JPDF in the form of the direct series in the products of
Hermite, for $\zeta$ and $J_1$, and Laguerre, for $q^2$
and $J_2$, polynomials:
\begin{widetext}
\begin{equation}
P_{\rm 2D}(\zeta, q^2, J_1, J_2) =  G_{\rm 2D} \left[ 1 + 
\sum_{n=3}^\infty \sum_{i,j,k,l=0}^{i+2 j+k+2 l=n} 
\frac{(-1)^{j+l}}{i!\;j!\; k!\; l!} 
\left\langle \zeta^i {q^2}^j {J_1}^k {J_2}^l \right\rangle_{\scriptscriptstyle{\mathrm{GC}}}
H_i\left(\zeta\right) L_j\left(q^2\right)
H_k\left(J_1\right) L_l\left(J_2\right)
\right]\,,
\label{eq:2DP_general}
\end{equation}
\end{widetext}
where $\sum_{i,j,k,l=0}^{i+2 j+k+2 l=n} $
stands for summation over all combinations
of non-negative $i,j,k,l$ such that $i+2j+k+2l$ adds 
to the order of the expansion term $n$. 
The coefficients of the expansion
are combinations of moments 
\begin{displaymath}
\left\langle \zeta^i {q^2}^j J_1^k J_2^l \right\rangle_{\scriptscriptstyle{\mathrm{GC}}} \!\! =
\frac{j! \; l!}{(-1)^{j+l}} 
\left\langle \vphantom{ \zeta^i {q^2}^j J_1^k J_2^l } \!
H_i\left(\zeta\right) L_j\left(q^2\right)
H_k\left(J_1\right) L_l\left(J_2\right) \!\right\rangle
\end{displaymath}
that vanish in the Gaussian limit.
In the lowest $n=3$ order they coincide with the cumulants
of our variables. 

In ND the Gaussian JPDF, 
$G_{\rm ND}(\zeta, q^2, J_1, J_{s\ge2})$, retains complete factorization
with respect to $\zeta, q^2, J_1$
\begin{equation}
G_{N\mathrm{D}}=
\frac{(\frac{N}{2})^{\frac{N}{2}} q^{N-2}}{2\pi \Gamma\left[\frac{N}{2}\right]} 
\exp\!\left[-\frac{\zeta^2}{2}\! -\! \frac{N q^2}{2}\! -\! \frac{J_1^2}{2}
\right]
{\mathcal G}(J_{s\ge2})
\label{eq:3DG}
\end{equation}
so in the moment expansion this sector always gives rise to Hermite
$H_i(\zeta)$, $H_k(J_1)$
and generalized Laguerre 
$L_j^{(N-2)/2}(N q^2/2)$ polynomials.
However, the distributions of the rest $J_{s\ge2}$ contained in
${\mathcal G}(J_{s \ge 2})$ are coupled. 

Specifically, in 3D, ${\mathcal G} (J_2,J_3) = \frac{25 \sqrt{5}}{6\sqrt{2\pi}} 
\exp[-\frac{5}{2} J_2]$ 
and $J_3$ is distributed uniformly between
$-J_2^{3/2}$ and $J_2^{3/2}$.
Let us denote the orthogonal polynomials in these two variables, 
$F_{lm}(J_2,J_3)$, where $l$ is the power of $J_2$ and $m$ is the power of 
$J_3$. They obey the orthonormality condition
\begin{equation}
\int_0^\infty  \!\!\!\! d J_2 \int_{-J_2^{3/2}}^{J_2^{3/2}} \!\! d J_3
{\mathcal G}\, F_{l m}(J_2,J_3) F_{l^\prime m^\prime}(J_2,J_3) 
= \delta_{l l^\prime} \delta_{m m^\prime} \nonumber
\end{equation}
We shall not give the full theory of these polynomials here, but note that
one can construct them by a Gram-Schmidt orthogonalization procedure
to any given order.  Two special cases 
$
F_{l0} = \sqrt{\frac{ 3 \times 2^l \times l!}{(3+2l)!!}} 
L_l^{(3/2)}\left(\frac{5}{2} J_2\right) 
$
and
$
F_{01} =  \frac{5}{\sqrt{21}} J_3
$
are sufficient to obtain the general expression for the Euler characteristic
of the excursion sets of the field to arbitrary order, and to calculate
the critical point and skeleton statistics to quartic order.

Hence, we write the moment expansion for invariant non-Gaussian JPDF,
$P_{\rm 3D}(\zeta, q^2, J_1, J_2, J_3)$ as a series in the power order of the
field, $n$ in the form
\begin{widetext}
\begin{eqnarray}
P_{\rm 3D} &=&  G_{\rm 3D}
\left[1+
\sum_{n=3}^\infty \sum_{i,j,k,l=0}^{i+2 j+k+2 l=n} 
\frac{(-1)^{j+l} 3^j 5^l \times 3}{i!\;(1+2j)!!\; k!\; (3+2l)!!} 
\left\langle \zeta^i {q^2}^j {J_1}^k {J_2}^l \right\rangle_{\scriptscriptstyle{\mathrm{GC}}}
H_i\left(\zeta\right) L_j^{(1/2)}\left(\frac{\scriptstyle 3}{\scriptstyle 2} q^2 \right)
H_k\left(J_1\right) L_l^{(3/2)}\left(\frac{\scriptstyle 5}{\scriptstyle 2} J_2\right) \right.
\nonumber \\
 && +
\sum_{n=3}^\infty \sum_{i,j,k=0}^{i + 2j + k + 3=n} 
\frac{(-1)^j 3^j \times 25}{i!\;(1+2j)!!\; k! \times 21} 
\left\langle \zeta^i {q^2}^j {J_1}^k J_3 \right\rangle_{\scriptscriptstyle{\mathrm{GC}}}
H_i\left(\zeta\right) L_j^{(1/2)}\left(\frac{\scriptstyle 3}{\scriptstyle 2} q^2 \right)
H_k\left(J_1\right) J_3
\nonumber \\
&&+ \left. 
\sum_{n=5}^\infty \sum_{i,j,k,l=0,m=1}^{i + 2j + k + 2l + 3m = n} 
\frac{(-1)^j 3^j c_{lm} }{i!\;(1+2j)!!\; k! } 
\left\langle \zeta^i {q^2}^j {J_1}^k {J_2}^l {J_3}^m \right\rangle_{\scriptscriptstyle{\mathrm{GC}}}
H_i\left(\zeta\right) L_j^{(1/2)}\left(\frac{\scriptstyle 3}{\scriptstyle 2} q^2 \right)
H_k\left(J_1\right) F_{lm}\left(J_2,J_3\right)
\right]\,.
\label{eq:3DP_general}
\end{eqnarray}
\end{widetext}
The first high order term contains $J_2$ but not $J_3$, the second is linear in $J_3$ and contains no $J_2$, while for where we left the normalization
coefficient $c_{lm}$ undetermined,
contains all the remaining combinations of
$J_2$ and $J_3$.
In this last term, $c_{01}=0$, so that the first
contribution to it, $l=m=1$, is of the fifth power of the field 
$\propto J_2 J_3$. The moment combinations that give the expansion coefficients are
found analogously to 2D case.


The formulas (\ref{eq:2DP_general}) and (\ref{eq:3DP_general}) 
provide the joint probability 
function of a field and its derivatives up to second order 
in terms of the invariant variables to an arbitrary order in the
moment expansion.
They allow to easily compute any rotation-invariant statistics that depend
exclusively on these descriptors of the field.
In particular, one can compute the Euler characteristic of the excursion sets
of the field analytically, the density of the
extrema, and the properties of the critical lines that describe the skeleton
of the field, as we describe in detail in the follow-up paper.

Determining  the topological Euler characteristic, $\chi$,
and the density of the extrema in the high excursion sets
$x > \nu$ of the random field 
are two related classical problems in the study of the geometry of 
a random field. Both are reduced \cite{Adler}
to the evaluation under the extremal condition $q^2 = 0$
of the statistical average of the Gaussian curvature of the
field surface given by the invariant $I_N$,
the difference being the various
conditions set on the  signs of the eigenvalues of the Hessian.
For the Euler characteristic integration is unconstrained
\begin{equation}
\frac{\chi(\nu)}{2}= (-1)^N
\int_\nu^\infty
\!\!\!\! d x \! \int \!\! dq^2 q^{N-1} \delta_D^N(q^2)\!
\int \prod_{s=1}^N d J_s P_{ND}\, I_N\,, \nonumber
\end{equation}
while to find the number density of extremal points of different types
one integrates over the regions in $J_s$ space 
where particular signature of the eigenvalues is maintained.
Since $I_s$ are just low order polynomials of the $J_s$ variables,
Eqs~(\ref{eq:2DP_general}) and (\ref{eq:3DP_general}) 
are well suited to perform the required calcualtions.

The Euler characteristic can be computed completely
by noting from Eq.~(\ref{eq:Js}) that $I_N$ depends only linearly
on $J_{s \ge 2}$ (e.g., 
$ I_3 = \frac{1}{27} \left({J_1}^3 - 3 J_1 J_2 + 2 J_3 \right)
$
in 3D), hence all terms in JPDF of higher order in $J_{s \ge 2}$
do not contribute.  The 2D and 3D results 
\footnote{We conjecture that Eq.~(\ref{eq:euler}) is valid for all $N$.}
can be combined in a very compact form if one re-expresses the 
coefficients
back in terms of the field $x$ itself and the invariants $I_s$
\begin{widetext}
\begin{eqnarray}
\lefteqn{\chi (\nu) = 
\frac{1}{2} \mathrm{Erfc} \left(\frac{\nu}{\sqrt{2}} \right)
\chi(-\infty) 
+ \frac{1}{\sqrt{2 \pi}} 
\exp\left(-\frac{\nu^2}{2}\right) \times
\frac{2}{(2 \pi)^{N/2}} \left(\frac{\gamma}{\sqrt{N}}\right)^N
\left[\vphantom{ \sum_{i,j,k=0}^{i+2 j+k=n} }
H_{N-1}(\nu) +
\right. } \nonumber \\
&& + \left. \sum_{n=3}^\infty  
\sum_{s=0}^N \gamma^{-s}
\sum_{i,j=0}^{i+2j=n-s}
\frac{(-N)^{j+s} (N-2) !! L_j^{(\frac{N-2}{2})}(0)}{i! (2j+N-2)!!} 
\left\langle x^i {q^2}^j I_s \right\rangle_{\scriptscriptstyle{\mathrm{GC}}} H_{i+N-s-1}(\nu)
\right]\,, \label{eq:euler}
\end{eqnarray}
\end{widetext}
where $i=0,s=N$ terms have been combined into the boundary term $\propto
\chi(-\infty)$ fixed by the topology of the manifold and should
be omitted from the sum. $I_0 \equiv 1$.

Calculation of the extrema number density produces an analytical result only in
2D. Even then the result to fourth order is already  too complicated 
to be reproduced here and is deferred to \cite{GP2}. Instead here we
demonstrate our results on two model non-Gaussian fields.

\begin{figure}
\center
\includegraphics[width=0.45\textwidth]{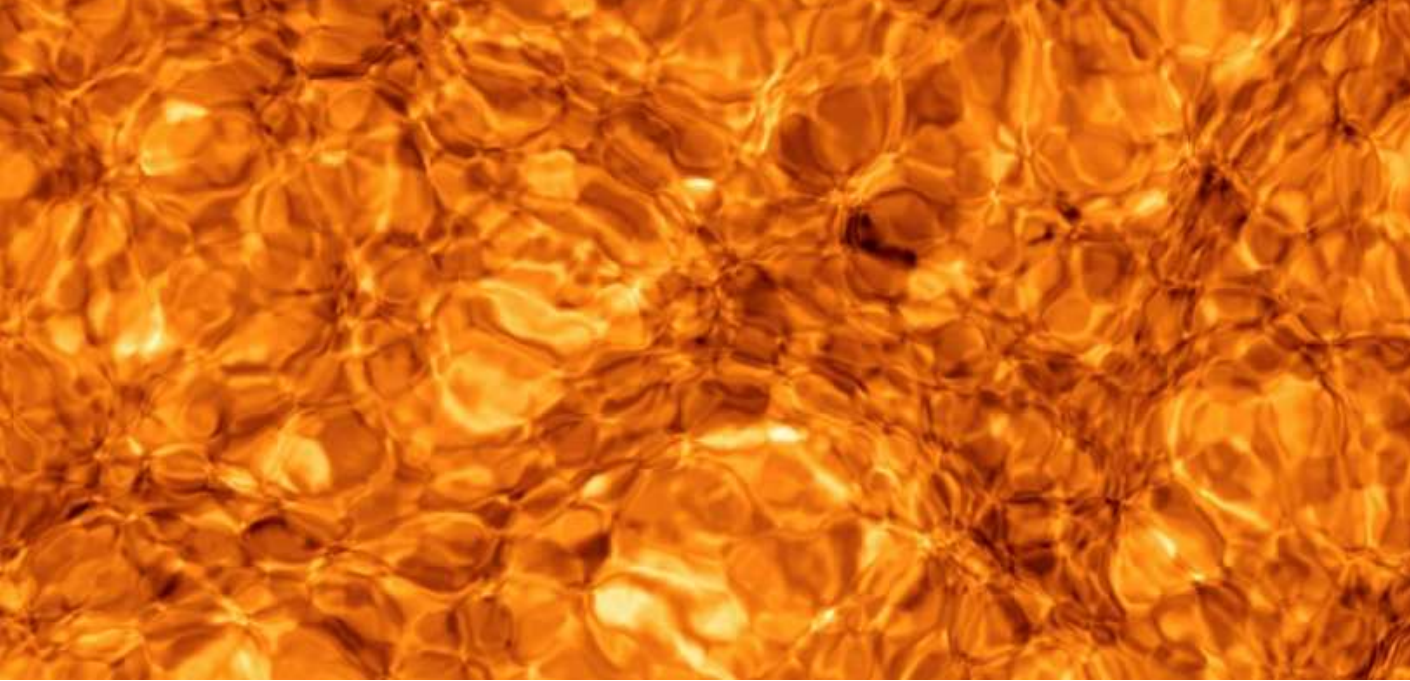}
\caption{A section of strongly non-Gaussian 2D field.}
\label{fig:field}
\vskip-0.45cm
\end{figure}

For the  a 2D case we generate
a Gaussian random field $x_G$ with the scale-invariant power spectrum 
$\propto k^{-1.5}$, 
smoothed with a Gaussian filter of $5$-pixel width. 
The underlying grid is then displaced by $ \alpha \nabla \Delta^{-1} x_G$ and
the initial field is resampled on this grid, producing a one-parameter
non-Gaussian field, $x^{(\alpha)}$. This toy model is able to produce high
level of non-Gaussianity without generating high field excursions, 
as demonstrated in Fig.~\ref{fig:field}
by a $2048^2$ realization of the field with $\alpha =0.5$.

In 3D, we use scale-invariant ($\propto k^{-1}$) simulations of
cosmological density evolved to the mildly non-linear
stage of gravitational instability
where the variance of inhomogeneities is $\sigma=0.1$ of the mean
density.

\begin{figure}
\vskip-1cm
\includegraphics[width=0.47\textwidth]{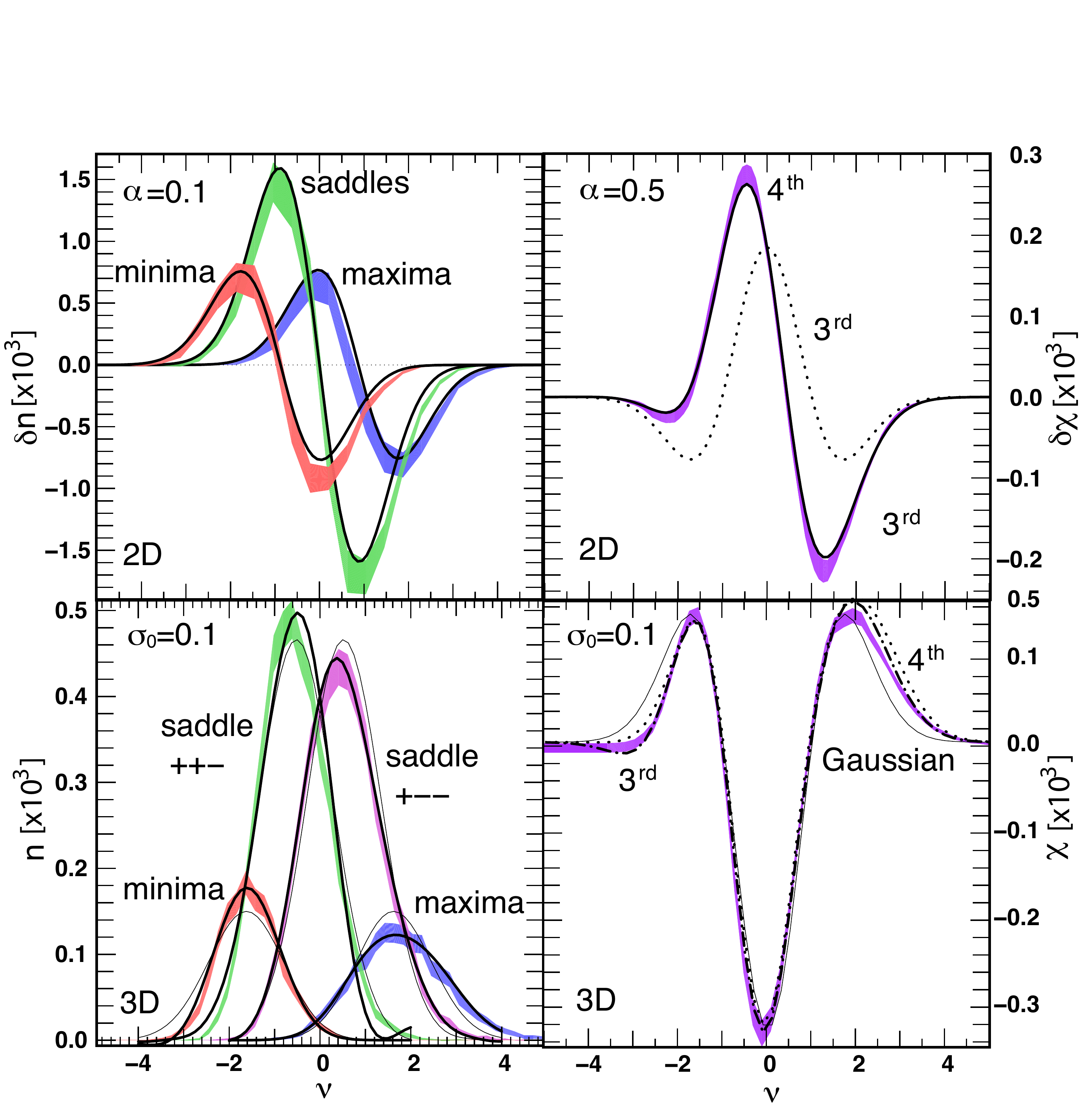}
\caption{The number of extrema ({\sl left}) and the Euler characteristic
({\sl right}) for the 2D toy model ({\sl top}) and 3D gravitational collapse
model ({\sl bottom}). In dimensional units
both quantities are given per $R_*^N$ volume defined as in \cite{pogoskel}.
The shaded bands correspond to $2\sigma$ variations over the mean measurements
of 100 (2D) or 25 (3D) realizations, while the curves give the $3^{\rm rd}$ and
$4^{\rm th}$ order predictions. 
A higher order correction is required to fit the strong non-Gaussian field shown in Fig.~\ref{fig:field}.
 In 2D only the correction  to the Gaussian limit is shown. In 3D,
the Gaussian prediction is shown as a thin line. 
} 
\label{fig:mainfig}
\vskip-0.3cm
\end{figure}

In Fig.~\ref{fig:mainfig} we plot the extrema count 
\footnote{Extrema counts are obtained 
by integrating analytically Eq.~(\ref{eq:2DP_general}) in 2D and
numerically Eq.~(\ref{eq:3DP_general}) in 3D over the relevant 
bounds of eigenvalues \cite{pogoskel}}
and the Euler characteristic computed within our formalism to 3rd and 4th order 
respectively, using the moments measured numerically in our toy
simulations. 
The {\sl left} column demonstrates the avenues 
our formalism opens for the theoretical study of critical points of
non-Gaussian fields. The {\sl right} column shows, on the example of the
Euler characteristic, the importance of accounting for high moments of
the expansion for some models of non-Gaussianity, as well as the convergence
process of the expansion for the gravitational instability models.

The ability to analyze the full series of the moment expansion for geometrical
statistics is important from several points of view.
In cosmological applications, e.g. a mildly non-linear gravitational
instability as simulated in 3D in Figure~\ref{fig:mainfig},
or lensing of CMB maps, one deals with perturbative deviation from Gaussianity governed
by the small value of the variance $\sigma$. 
In this case Gram-Charlier series can be rearranged 
in (asymptotic) Edgeworth \cite{blin} power series in $\sigma$ \footnote{It is worth noting that leading $n=3$ polynomial term coincides with linear in $\sigma$ correction
to PDF.}
and the full expansion can be used
to obtain an increasingly accurate description of the statistics, and conduct
estimates of residual errors. 
Outside of established cosmological perturbation theory,
theoretical formalism that deals globally with all the terms is all the  more important.
It may allow us to design a customized truncation criteria, which may not
be uniform across all the statistics, or even for all the partial contributions
to a given statistical descriptor \footnote{e.g. 
 Eq.~(\ref{eq:euler}) can be rearranged as series in the Hermite
polynomials order, where coefficients are (asymptotic) series
in the moments of the field. Convergence properties and the best truncation
may differ for each coefficient.}.

In the cosmological context, non-Gaussianity of extragalactic fields
can be used to constrain the dark energy equation of state via
3D galactic surveys, or shed  light on the physics of the early Universe
through 2D CMB maps.
The predictions of theoretical models are often given
through the hierarchy of the differences between the
moments to their Gaussian limit. Higher order moments 
are generally difficult to test directly in real-life observations,
due to their sensitivity to very rare events. The geometrical
analysis of the critical events in the field provides more robust measures
of non-Gaussianity and is becoming an active 
field of investigation \cite{Gott,Park}.
The present paper advances the formal development of this subject, 
opening ways to expand the list and sharpen statistical tools that
may detect unique signatures of fundamental processes in our Universe.

\bibliography{dofzlet}

\begin{thebibliography}{16}
\expandafter\ifx\csname natexlab\endcsname\relax\def\natexlab#1{#1}\fi
\expandafter\ifx\csname bibnamefont\endcsname\relax
  \def\bibnamefont#1{#1}\fi
\expandafter\ifx\csname bibfnamefont\endcsname\relax
  \def\bibfnamefont#1{#1}\fi
\expandafter\ifx\csname citenamefont\endcsname\relax
  \def\citenamefont#1{#1}\fi
\expandafter\ifx\csname url\endcsname\relax
  \def\url#1{\texttt{#1}}\fi
\expandafter\ifx\csname urlprefix\endcsname\relax\def\urlprefix{URL }\fi
\providecommand{\bibinfo}[2]{#2}
\providecommand{\eprint}[2][]{\url{#2}}

\bibitem[{\citenamefont{{Adler}}(1981)}]{Adler}
\bibinfo{author}{\bibfnamefont{R.~J.} \bibnamefont{{Adler}}},
  \emph{\bibinfo{title}{The Geometry of Random Fields}}
  (\bibinfo{publisher}{The Geometry of Random Fields, Chichester: Wiley},
  \bibinfo{year}{1981}).

\bibitem[{\citenamefont{{Doroshkevich}}(1970)}]{Doroshkevich}
\bibinfo{author}{\bibfnamefont{A.~G.} \bibnamefont{{Doroshkevich}}},
  \bibinfo{journal}{Astrofizika} \textbf{\bibinfo{volume}{6}},
  \bibinfo{pages}{581} (\bibinfo{year}{1970}).

\bibitem[{\citenamefont{{Bardeen} et~al.}(1986)\citenamefont{{Bardeen}, {Bond},
  {Kaiser}, and {Szalay}}}]{BBKS}
\bibinfo{author}{\bibfnamefont{J.~M.} \bibnamefont{{Bardeen}}},
  \bibinfo{author}{\bibfnamefont{J.~R.} \bibnamefont{{Bond}}},
  \bibinfo{author}{\bibfnamefont{N.}~\bibnamefont{{Kaiser}}}, \bibnamefont{and}
  \bibinfo{author}{\bibfnamefont{A.~S.} \bibnamefont{{Szalay}}},
  \bibinfo{journal}{\apj} \textbf{\bibinfo{volume}{304}}, \bibinfo{pages}{15}
  (\bibinfo{year}{1986}).

\bibitem[{\citenamefont{{Bond} et~al.}(1996)\citenamefont{{Bond}, {Kofman}, and
  {Pogosyan}}}]{bkp}
\bibinfo{author}{\bibfnamefont{J.~R.} \bibnamefont{{Bond}}},
  \bibinfo{author}{\bibfnamefont{L.}~\bibnamefont{{Kofman}}}, \bibnamefont{and}
  \bibinfo{author}{\bibfnamefont{D.}~\bibnamefont{{Pogosyan}}},
  \bibinfo{journal}{\nat} \textbf{\bibinfo{volume}{380}}, \bibinfo{pages}{603}
  (\bibinfo{year}{1996}), \eprint{arXiv:astro-ph/9512141}.

\bibitem[{\citenamefont{{Gott}}(1988)}]{Gott1988}
\bibinfo{author}{\bibfnamefont{J.~R.~I.} \bibnamefont{{Gott}}},
  \bibinfo{journal}{\pasp} \textbf{\bibinfo{volume}{100}},
  \bibinfo{pages}{1307} (\bibinfo{year}{1988}).

\bibitem[{\citenamefont{{Matsubara}}(1994)}]{Matsubara0}
\bibinfo{author}{\bibfnamefont{T.}~\bibnamefont{{Matsubara}}},
  \bibinfo{journal}{\apjl} \textbf{\bibinfo{volume}{434}}, \bibinfo{pages}{L43}
  (\bibinfo{year}{1994}).

\bibitem[{\citenamefont{{Matsubara}}(2003)}]{Matsubara}
\bibinfo{author}{\bibfnamefont{T.}~\bibnamefont{{Matsubara}}},
  \bibinfo{journal}{\apj} \textbf{\bibinfo{volume}{584}}, \bibinfo{pages}{1}
  (\bibinfo{year}{2003}).

\bibitem[{\citenamefont{{Colombi} et~al.}(2000)\citenamefont{{Colombi},
  {Pogosyan}, and {Souradeep}}}]{2000PhRvL..85.5515C}
\bibinfo{author}{\bibfnamefont{S.}~\bibnamefont{{Colombi}}},
  \bibinfo{author}{\bibfnamefont{D.}~\bibnamefont{{Pogosyan}}},
  \bibnamefont{and}
  \bibinfo{author}{\bibfnamefont{T.}~\bibnamefont{{Souradeep}}},
  \bibinfo{journal}{Physical Review Letters} \textbf{\bibinfo{volume}{85}},
  \bibinfo{pages}{5515} (\bibinfo{year}{2000}),
  \eprint{arXiv:astro-ph/0011293}.

\bibitem[{\citenamefont{{Mecke} et~al.}(1994)\citenamefont{{Mecke}, {Buchert},
  and {Wagner}}}]{Buchert}
\bibinfo{author}{\bibfnamefont{K.~R.} \bibnamefont{{Mecke}}},
  \bibinfo{author}{\bibfnamefont{T.}~\bibnamefont{{Buchert}}},
  \bibnamefont{and} \bibinfo{author}{\bibfnamefont{H.}~\bibnamefont{{Wagner}}},
  \bibinfo{journal}{\aap} \textbf{\bibinfo{volume}{288}}, \bibinfo{pages}{697}
  (\bibinfo{year}{1994}), \eprint{arXiv:astro-ph/9312028}.

\bibitem[{\citenamefont{{Novikov} et~al.}(2006)\citenamefont{{Novikov},
  {Colombi}, and {Dor{\'e}}}}]{NCD}
\bibinfo{author}{\bibfnamefont{D.}~\bibnamefont{{Novikov}}},
  \bibinfo{author}{\bibfnamefont{S.}~\bibnamefont{{Colombi}}},
  \bibnamefont{and}
  \bibinfo{author}{\bibfnamefont{O.}~\bibnamefont{{Dor{\'e}}}},
  \bibinfo{journal}{\mnras} \textbf{\bibinfo{volume}{366}},
  \bibinfo{pages}{1201} (\bibinfo{year}{2006}),
  \eprint{arXiv:astro-ph/0307003}.

\bibitem[{\citenamefont{{Pogosyan} et~al.}(2009)\citenamefont{{Pogosyan},
  {Pichon}, {Gay}, {Prunet}, {Cardoso}, {Sousbie}, and {Colombi}}}]{pogoskel}
\bibinfo{author}{\bibfnamefont{D.}~\bibnamefont{{Pogosyan}}},
  \bibinfo{author}{\bibfnamefont{C.}~\bibnamefont{{Pichon}}},
  \bibinfo{author}{\bibfnamefont{C.}~\bibnamefont{{Gay}}},
  \bibinfo{author}{\bibfnamefont{S.}~\bibnamefont{{Prunet}}},
  \bibinfo{author}{\bibfnamefont{J.~F.} \bibnamefont{{Cardoso}}},
  \bibinfo{author}{\bibfnamefont{T.}~\bibnamefont{{Sousbie}}},
  \bibnamefont{and}
  \bibinfo{author}{\bibfnamefont{S.}~\bibnamefont{{Colombi}}},
  \bibinfo{journal}{\mnras} \textbf{\bibinfo{volume}{396}},
  \bibinfo{pages}{635} (\bibinfo{year}{2009}), \eprint{arXiv:0811.1530}.

\bibitem[{\citenamefont{Chambers}(1967)}]{Chambers}
\bibinfo{author}{\bibfnamefont{J.~M.} \bibnamefont{Chambers}},
  \bibinfo{journal}{Biometrika} \textbf{\bibinfo{volume}{54}},
  \bibinfo{pages}{367} (\bibinfo{year}{1967}).

\bibitem[{\citenamefont{{Gay} et~al.}(2010)\citenamefont{{Gay}, {Pichon}, and
  {Pogosyan}}}]{GP2}
\bibinfo{author}{\bibfnamefont{C.}~\bibnamefont{{Gay}}},
  \bibinfo{author}{\bibfnamefont{C.}~\bibnamefont{{Pichon}}}, \bibnamefont{and}
  \bibinfo{author}{\bibfnamefont{D.}~\bibnamefont{{Pogosyan}}}
  (\bibinfo{year}{2010}), \eprint{in prep.}

\bibitem[{\citenamefont{{Blinnikov} and {Moessner}}(1998)}]{blin}
\bibinfo{author}{\bibfnamefont{S.}~\bibnamefont{{Blinnikov}}} \bibnamefont{and}
  \bibinfo{author}{\bibfnamefont{R.}~\bibnamefont{{Moessner}}},
  \bibinfo{journal}{{\aap}{\it S}} \textbf{\bibinfo{volume}{130}},
  \bibinfo{pages}{193} (\bibinfo{year}{1998}), \eprint{arXiv:astro-ph/9711239}.

\bibitem[{\citenamefont{{Gott} et~al.}(2009)\citenamefont{{Gott}, {Choi},
  {Park}, and {Kim}}}]{Gott}
\bibinfo{author}{\bibfnamefont{J.~R.} \bibnamefont{{Gott}}},
  \bibinfo{author}{\bibfnamefont{Y.-Y.} \bibnamefont{{Choi}}},
  \bibinfo{author}{\bibfnamefont{C.}~\bibnamefont{{Park}}}, \bibnamefont{and}
  \bibinfo{author}{\bibfnamefont{J.}~\bibnamefont{{Kim}}},
  \bibinfo{journal}{\apjl} \textbf{\bibinfo{volume}{695}}, \bibinfo{pages}{L45}
  (\bibinfo{year}{2009}), \eprint{0812.1406}.

\bibitem[{\citenamefont{{Park} et~al.}(2005)\citenamefont{{Park}, {Choi},
  {Vogeley}, {Gott}, {Kim}, {Hikage}, {Matsubara}, {Park}, {Suto}, and
  {Weinberg}}}]{Park}
\bibinfo{author}{\bibfnamefont{C.}~\bibnamefont{{Park}}},
  \bibinfo{author}{\bibfnamefont{Y.-Y.} \bibnamefont{{Choi}}},
  \bibinfo{author}{\bibfnamefont{M.~S.} \bibnamefont{{Vogeley}}},
  \bibinfo{author}{\bibfnamefont{J.~R.~I.} \bibnamefont{{Gott}}},
  \bibinfo{author}{\bibfnamefont{J.}~\bibnamefont{{Kim}}},
  \bibinfo{author}{\bibfnamefont{C.}~\bibnamefont{{Hikage}}},
  \bibinfo{author}{\bibfnamefont{T.}~\bibnamefont{{Matsubara}}},
  \bibinfo{author}{\bibfnamefont{M.-G.} \bibnamefont{{Park}}},
  \bibinfo{author}{\bibfnamefont{Y.}~\bibnamefont{{Suto}}}, \bibnamefont{and}
  \bibinfo{author}{\bibfnamefont{D.~H.} \bibnamefont{{Weinberg}}},
  \bibinfo{journal}{\apj} \textbf{\bibinfo{volume}{633}}, \bibinfo{pages}{11}
  (\bibinfo{year}{2005}), \eprint{arXiv:astro-ph/0507059}.

\end{thebibliography}
\end{document}